\documentclass{aastex631}

\usepackage{natbib}
\usepackage{amsmath}

\begin{document}

\title{The Impact of Planetary Phase Functions on Exo-Earth Detectability with EXOSIMS}
\author[0000-0001-5477-8588]{Searra Foote}
\affiliation{Lunar \& Planetary Laboratory, University of Arizona, Tucson, AZ 85721, USA}
\affiliation{Habitability, Atmospheres, and Biosignatures Laboratory, University of Arizona, Tucson, AZ 85721, USA}

\author[0000-0002-3196-414X]{Tyler D. Robinson}
\affiliation{Lunar \& Planetary Laboratory, University of Arizona, Tucson, AZ 85721 USA}
\affiliation{Habitability, Atmospheres, and Biosignatures Laboratory, University of Arizona, Tucson, AZ 85721, USA}
\affiliation{NASA Nexus for Exoplanet System Science Virtual Planetary Laboratory, University of Washington, Box 351580, Seattle, WA 98195, USA}

\author[0000-0002-4852-6330]{Rhonda Morgan}
\affiliation{Jet Propulsion Laboratory, California Institute of Technology, 4800 Oak Grove Dr., Pasadena, CA, 91109, USA}

\author[0000-0002-8711-7206]{Dmitry Savransky}
\affiliation{Sibley School of Mechanical and Aerospace Engineering, Cornell University, Ithaca, NY, 14853, USA}
\affiliation{Carl Sagan Institute, Cornell University, Ithaca, NY, 14853, USA}

\author[0000-0002-1830-8260]{Mario Damiano}
\affiliation{Jet Propulsion Laboratory, California Institute of Technology, 4800 Oak Grove Dr., Pasadena, CA, 91109, USA}

\author{Armen Tokadjian}
\affiliation{Jet Propulsion Laboratory, California Institute of Technology, 4800 Oak Grove Dr., Pasadena, CA, 91109, USA}

\begin{abstract}

The under-development NASA Habitable Worlds Observatory (HWO) aims to provide breakthroughs in exoplanet science, yet the most effective approaches to modeling the detection and characterization of potentially Earth-like worlds with HWO remain uncertain despite being essential considerations for mission design. In this work, we aim to better model and understand detection metrics through the use of EXOSIMS (Exoplanet Open-Source Imaging Mission Simulator), an exoplanet yield modeling tool. Yield modeling requires representing planetary brightness via a planetary phase curve. Earth’s true visual phase curve is non-Lambertian, deviating from the idealized Lambertian model in EXOSIMS, particularly at phase angles beyond 90 degrees(i.e., quadrature). This leads to underestimating Earth's brightness. To address this, we incorporate phase-dependent reflectance from a high-fidelity Earth model into EXOSIMS for physically motivated simulations. We explore and quantify differences in phase-dependent detections, finding that the realistic Earth phase function produces modest changes in the median number of detected exo-Earths and systematically redistributes detections towards medium to high phase angles where Earth is intrinsically brighter than a Lambertian approximation. Additionally, we explore the role of coronagraph inner working angle (IWA) by running simulations across multiple IWA values with both phase functions, revealing that smaller IWAs expand access to a broader range of orbital phases, altering the resulting phase-angle distribution of detections. Together, these results demonstrate that realistic phase functions and IWA parameters both have measurable impacts on yield estimates for an HWO-like mission and highlight the need to more realistically represent Earth-like worlds in yield modeling.

\end{abstract}


\section{Introduction} \label{sec:intro}

The National Academies’ ``Decadal Survey on Astronomy and Astrophysics 2020'' \citep[Astro2020;][]{Astro2020} set forth a vision for the coming decade that would take key steps towards a long-awaited goal of astronomy and astrophysics: the direct imaging of Earth-sized planets in the habitable zones of nearby, Sun-like stars. The Habitable Worlds Observatory (HWO) is a response to this Astro2020 recommendation that NASA build a large space-telescope capable of directly imaging Earth-like exoplanets. While HWO is set to launch in the 2040’s, design for this mission is happening now.

The successful development of a mission like HWO requires tools and science that connect mission goals to telescope architecture using yield models to achieve these goals \citep{Stark2024, Morgan2017, Delacroix2016}. Yield models are computational tools used to estimate the expected science return of a  given observational mission. These models incorporate mission-specific parameters, such as telescope aperture size, inner working angle (IWA), and throughput, along with astrophysical inputs such as planetary occurrence rates and orbital distributions. In the case of direct imaging missions, yield models also account for the brightness of the planets relative to their host stars, expressed as the planet-to-star flux ratio \citep{brown2005single}.

Tools such as the Exoplanet Open-Source Imaging Mission Simulator (EXOSIMS) \citep{Savransky2015} and the Altruistic Yield Optimization (AYO) \citep{Stark2014} have been pivotal in advancing the field of direct imaging yield calculations. For example, EXOSIMS allows for robust simulations of exoplanet detection scenarios, offering insights into detection probabilities across various mission architectures \citep{Morgan2017}. The AYO tool, meanwhile, focuses on allocating time per target to maximize the accumulated completeness for specific science objectives. Together, these yield models have helped refine mission designs and improve our understanding of what is achievable with future observatories.

An essential aspect of yield modeling is accurately representing phase-dependent brightness, which depends, in part, on the planetary phase function \citep{sobolev1975light}. A phase function describes how the reflected-light brightness of an astronomical object, such as a planet, varies with its phase angle \citep{seager2010exoplanet}. Understanding phase functions is crucial in exoplanet studies, as they provide insights into how light interacts with a planet's surface and atmosphere, influencing its observed brightness from Earth. Phase functions are essential for simulating the detectability of exoplanets in observational studies \citep{Madhusudhan2012}. By accurately modeling how planets reflect light at different angles, we can better predict which planets are more likely to be detected based on their brightness, which varies with the observer's position relative to the planet and its star.

The detectability of exoplanets is tied to their changing illumination and brightness as seen from Earth. A key quantity is the previously-mentioned phase-dependent planet-to-star flux ratio, given by,
\begin{equation}
\frac{F_{\rm p}}{F_{\rm s}} = A_{\rm g} \Phi(\alpha) \left( \frac{R_{\rm p}}{a} \right)^2 \ ,
\label{eq:flux_ratio}
\end{equation}
where $A_{\rm g}$ is the wavelength-dependent planetary geometric (i.e., full phase) albedo, $\Phi$ is the wavelength- and phase-dependent planetary phase function (normalized to unity at full phase), $\alpha$ is the phase angle (i.e., the star-planet-observer angle), $R_{\rm p}$ is the planetary radius, and $a$ is the planetary orbital distance \citep{sudarsky2005phase, cahoy2010exoplanet}. This expression highlights that the flux ratio depends not only on the intrinsic properties of the planet, such as albedo and radius, but also on its position in orbit and the resulting phase angle. Yield tools like EXOSIMS and AYO generally incorporate a realistic geometric albedo (especially for Earth-like targets) and realistic size and orbital parameters. However, these tools usually adopt a Lambert phase function. 

The Lambert phase function ($\Phi_{\rm L}$) is a simplified model used to describe the reflective properties of a globe assuming it is an isotropic reflector. This function is based on Lambert's cosine law, which states that the intensity of light reflected off a surface is proportional to the cosine of the angle between the incident light and the surface normal. The Lambert phase function can be expressed as,
\begin{equation}
\Phi_{\rm L}(\alpha) = \frac{\sin(\alpha) + (\pi - \alpha) \cos(\alpha)}{\pi} \ .    
\end{equation}
The Lambert phase function is widely adopted in yield calculations due to its simplicity and ease of use as a baseline model \citep{Luger2022}. It assumes isotropic scattering at each location on the planetary disk, making it a useful approximation for some planetary surfaces. However, it does not capture the complexities of real planetary atmospheres or surfaces, which typically exhibit anisotropic, wavelength-dependent scattering \citep{Seager2018}.

Yield tools like EXOSIMS and others often default to the Lambert phase function because it provides a straightforward way to estimate the expected brightness of exoplanets across all phase angles. Recent studies \citep[e.g.,][]{mayorga2016jupiters,mayorga2020reflected, garcia2017titan, cooper2025extreme} have shown that the Lambertian assumption fails to capture the complex phase behavior observed in real planets, reinforcing the need for more realistic phase function models in yield simulations. This approach sets a consistent standard for comparison when exploring the impact of more complex or realistic planetary reflectance models. Beyond HWO, the improvements in phase function treatment are applicable to a wide range of direct imaging missions where realistic phase-dependent brightness models are expected to similarly affect yield estimates and observational strategies.

Earth, the primary target in many yield calculations, exemplifies why the Lambert phase function can fall short. Observational data reveals that Earth's phase function is markedly non-Lambertian \citep{Qiu2003,Palle, robinson2026}. At crescent phases, Earth is significantly brighter than the Lambert model predicts, largely due to specular reflection from water surfaces and cloud forward scattering \citep{robinson2026}. Therefore, Earth is an ideal case study for understanding the limitations of the Lambert approximation and for developing more accurate phase functions. Forward models, such as those developed by \citet{Robinson2010}, accurately reproduce the phase behavior observed for Earth. 

A realistic phase function for Earth, derived from observations and high-fidelity models, provides a more precise representation of Earth-like exoplanets' brightness variations across different phase angles \citep{Robinson2010}. Accurate modeling of the phase function, particularly for Earth-like planets, is critical, as the combination of planetary brightness and apparent separation from a host star both influence the likelihood of detection. Thus, the incorporation of realistic phase curves into yield tools would help to ensure that mission planning accounts for the relation between planetary properties, orbital configurations, and observational constraints. However, there is still work that needs to be done to ensure yields are as accurate as possible. Below, we continue the community efforts to develop accurate yield tools begin by updating the Earth phase function in EXOSIMS with realistic data. 
 
Observability, in turn, relies critically on the inner working angle (IWA), a key instrument parameter for a direct imaging mission. The IWA sets the smallest angular separation at which a planet can be detected relative to its host star. While a coronagraph must suppress starlight to high contrast levels, regions very close to the star remain inaccessible due to diffraction and instrument limitations. Planets located at small star-planet separations or small phase angles might not be observable even if they are bright enough to detect in principle. For an HWO architecture, the IWA determines which orbital phases of an Earth twin can be accessed 

\section{Methods} \label{sec:methods}
\subsection{EXOSIMS Overview}

EXOSIMS is a versatile and robust simulation tool for designing and optimizing space-based exoplanet direct imaging missions \citep{Savransky2015,Morgan2017}. Its main purpose is to evaluate the feasibility and expected scientific yield of missions aimed at detecting and characterizing exoplanets, especially Earth-like worlds. Built in a modular Python framework, EXOSIMS incorporates astrophysical modeling, observatory capabilities, and operational constraints to generate realistic simulations of mission performance. 

The modular nature of EXOSIMS allows users to focus on specific aspects of the simulation \citep{Delacroix2016}. Each module represents a discrete component of the mission pipeline: the astrophysical population, the observatory’s technical parameters, the observation scheduling strategy, and the detection and characterization processes. Users can swap or modify individual modules to test parameters, enabling the customization of simulations to address unique scientific or engineering questions. EXOSIMS uses a JSON script to define these input parameters for its simulations, providing a flexible and structured method to configure these mission scenarios. By parsing this input file, EXOSIMS initializes the simulation environment and ensures that the specified parameters govern the Monte Carlo mission simulations while EXOSIMS is running.  

\begin{table}[ht!]
\centering
\caption{Example subset of EXOSIMS parameters relevant to mission architecture and simulation configuration used in this work provided through a JSON script.}
\begin{tabular}{|c|c|}
\hline
\textbf{Parameter} & \textbf{Value}             \\ \hline
\texttt{missionLife}        & 5 years                          \\ \hline
\texttt{pupilDiam}          & 6 meters                       \\ \hline
\texttt{IWA}            & 0.054 arcseconds ($\sim$3\,$\lambda/D$)                 \\ \hline
\texttt{OWA}                & 0.422 arcseconds (26.5\,$\lambda/D$)                       \\ \hline
\texttt{PlanetPopulation}   & AlbedoByRadiusDulzPlavchan \\ \hline
\texttt{StarCatalog}        & EXOCAT1                    \\ \hline
\texttt{Completeness}       & BrownCompleteness          \\ \hline
\texttt{SimulatedUniverse}  & DulzPlavchanUniverseEarthsOnly              \\ \hline
\texttt{SurveySimulation}   & coroOnlyScheduler          \\ \hline
\end{tabular}

\label{tbl:instrument_params}
\end{table}

Table~\ref{tbl:instrument_params} highlights a subset of important parameters that are included in a JSON script for simulations presented here as a baseline setup. A comprehensive description of the EXOSIMS framework, module, and mission parameters is provided in \citep{Savransky2015,Morgan2017} EXOXIMS is comprised of modular components The \texttt{missionLife} specifies the duration of the mission in years, setting the operational timeframe for data collection, while the \texttt{pupilDiam} indicates the diameter of the telescope’s primary aperture in meters, which determines its light-gathering power and angular resolution. The inner working angle (\texttt{IWA}) and outer working angle (\texttt{OWA}), measured in arcseconds, define the smallest and largest angular separations at which the telescope can discern exoplanets from their host stars. The \texttt{PlanetPopulation} parameter identifies the planetary population model, here using the Dulz-Plavchan model, which simulates planet properties such as radius and albedo \citep{dulz2019exoplanet}. The \texttt{StarCatalog}, specified as \texttt{EXOCAT1} \citep{turnbull2015exocat}, provides the list of stars targeted for observations , while the \texttt{Completeness} module (\texttt{BrownCompleteness}) estimates the likelihood of detecting planets based on observational constraints \citep{brown2005single}. The \texttt{SimulatedUniverse} parameter, set to \texttt{DulzPlavchanUniverseEarthsOnly}, generates a synthetic planetary system containing only Earth-like planets for the simulation. Finally, the \texttt{SurveySimulation}, using \texttt{coroOnlyScheduler}, dictates the scheduling algorithm for prioritizing and sequencing observations focused exclusively on coronagraph science. Together, these parameters shape the simulated mission's performance and scientific yield.

The simulation process in EXOSIMS starts with the generation of a synthetic universe. This involves creating a population of stars, each then assigned a modeled planetary system. The stars are drawn from a real input catalog containing properties relevant to exoplanet imaging, such as luminosity, distance, and angular separation from other stars. Planetary systems are then assigned to each star using statistical distributions for parameters such as orbital elements, planet radii, masses, and albedos. These distributions are derived from observational data and theoretical models, ensuring the simulated population is as realistic as possible. Here, to isolate the impact of phase function assumptions on yield, we model all Earth-sized planets in the habitable zone as Earth twins within EXOSIMS.

The next step in EXOSIMS involves modeling the observatory and its instruments. This includes detailed specifications of the telescope, such as its aperture size, spectral bandpasses, and sensitivity limits. The performance of the instrument, particularly its ability to distinguish planets from the glint of their host stars, is modeled using parameters such as inner working angle (IWA), contrast limits, and throughput. EXOSIMS also incorporates observatory constraints, such as field of view, slew rates, and star accessibility, to simulate operational limitations as realistically as possible.

Once the astrophysical model and observatory parameters are set, EXOSIMS simulates the observation process. This is done through the generation of a complete end-to-end simulation of an observing program. Observation scheduling algorithms prioritize stars based on factors including their completeness and revisit strategies for characterizations. For each observation, EXOSIMS evaluates the observability of a planet by comparing its apparent brightness and projected separation to the instrument’s capabilities. These calculations account for the geometry of the planetary system, including the planet’s position in its orbit at the time of observation, and are influenced by assumptions about phase functions and albedo. The observing program evolves as the mission progresses. EXOSIMS completes these tasks through Monte Carlo mission simulations to conduct iterative, probabilistic analyses. By sampling from distributions of planetary and stellar properties, EXOSIMS generates synthetic populations and evaluates their detectability across ensembles of mission simulations. Within a single ensemble, the mission architecture and observing strategy are held constant while the random draw of planets varies. Thus, multiple ensembles are needed to consider the impacts of changes to the observatory or instrument. Each iteration computes quantities such as integration time required to achieve prescribed signal-to-noise ratio (SNR) thresholds and phase-dependent brightness, allowing EXOSIMS to statistically predict mission yield.

The outputs of EXOSIMS are highly detailed and provide insights into the performance of the simulated mission. These outputs include the number of detected planets, their orbital parameters, the characteristics of characterized planets, and the time spent observing each target. By comparing outputs across multiple ensembles with varied mission parameters, users can assess how changes in mission architecture affect progress toward desired science goals. For example, they can assess how changing the telescope's aperture or improving the IWA may affect the detection of Earth-like planets in the habitable zone.

EXOSIMS also tracks detection metrics such as the working angle and calculates derived values like the separation at the time of detection. This separation can be normalized by the semimajor axis to provide a dimensionless metric for comparing detections across different planetary systems. By simulating detections across multiple random seeds, corresponding to different random realizations of planetary systems, EXOSIMS samples the stochastic variability of the simulated population and reduces sensitivity to any single random realization.

In addition to simulating detections, EXOSIMS models the process of planetary characterization. Once a planet is detected, the mission must allocate additional time to obtain spectral or photometric measurements, which help determine the planet’s atmospheric composition, temperature, and potential habitability. EXOSIMS allows users to define flexible mission operating rules that incorporate the trade-offs between detecting new planets and characterizing known ones. By configuring observation priorities and scheduling strategies, users can explore the relative weightings of surveying for new planets versus revisiting previously detected targets, helping to optimize the balance between these competing objectives.

In EXOSIMS, the planet–star flux ratio is computed using Equation~\ref{eq:flux_ratio} above, where the phase function, $\phi \left( \alpha \right)$, modulates the reflected light as a function of phase angle. For this work, we assume that all Earth-sized planets are ``Earth twins,'' which can be defined as sharing the same geometric albedo and phase behavior as modern Earth in order to isolate the effects of differing phase functions. Although this is a simplification, it enables a controlled comparison between a simple Lambertian model and an observationally informed phase function.

\subsection{Phase Function Data}

Earth phase function data was sourced from high-fidelity simulations presented in \citet{Robinson2010}, which used the Virtual Planetary Laboratory Three-Dimensional Spectral Earth Model \citep{Tinetti2006,Robinson2011,Schwieterman2015}. This model, which has been extensively validated against phase-dependent Earth observations, captures the wavelength-dependent, non-Lambertian features of Earth’s brightness variations across phase angles, including the significantly higher brightness observed at crescent phases due to contributions from atmospheric scattering, cloud coverage, and specular reflection from surface water. The adopted simulation is through an entire calendar year for an observer situated in Earth's orbital plane. Thus, the observer sees the entire range of phase angles twice (i.e., waxing and waning). Furthermore, the orbit is situated so that secondary eclipse occurs at vernal equinox, so that there are Northern summer- and winter-aligned branches of the phase curve. However, phase-angle effects dominate over seasonal variations in the phase curve.

\begin{figure}[ht!]
    \centering
    \includegraphics[width=0.5\linewidth]{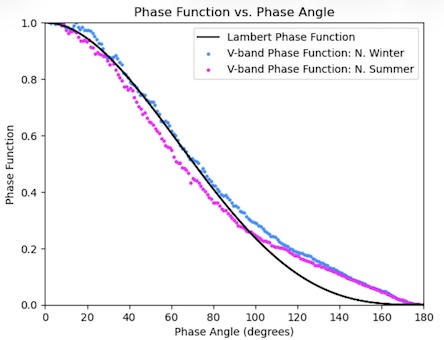}
    \caption{Comparison of the Lambert phase function (black) with the season-dependent, V-band--integrated phase functions for Earth. The V-band phase functions account for brightness variations due to Northern winter (blue) and Northern summer (pink) conditions, demonstrating enhanced brightness at crescent phase angles compared to the Lambert model. These band-integrated phase curves are used in the EXOSIMS simulations presented in this work.}
    \label{fig:phasefunc}
\end{figure}

Figure~\ref{fig:phasefunc} shows a comparison between the season-dependent, realistic Earth phase function from \citet{Robinson2010} and a Lambert phase function. The Lambert phase function assumes uniform surface scattering, while the realistic phase function incorporates Earth’s atmospheric scattering, direction-dependent cloud scattering, and specular reflection from surface water. The realistic Earth phase function can be more than twice as large as the Lambert phase function at some crescent phase angles, potentially corresponding to more than a fourfold decrease in requisite exposure time when compared to the Lambert model.

\begin{figure}[ht!]
    \centering
    \includegraphics[width=0.6\textwidth]{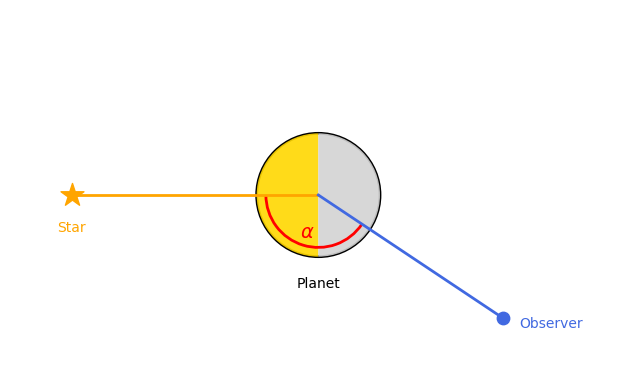}

    \hspace{0.05\textwidth}
    \includegraphics[width=0.6\textwidth]{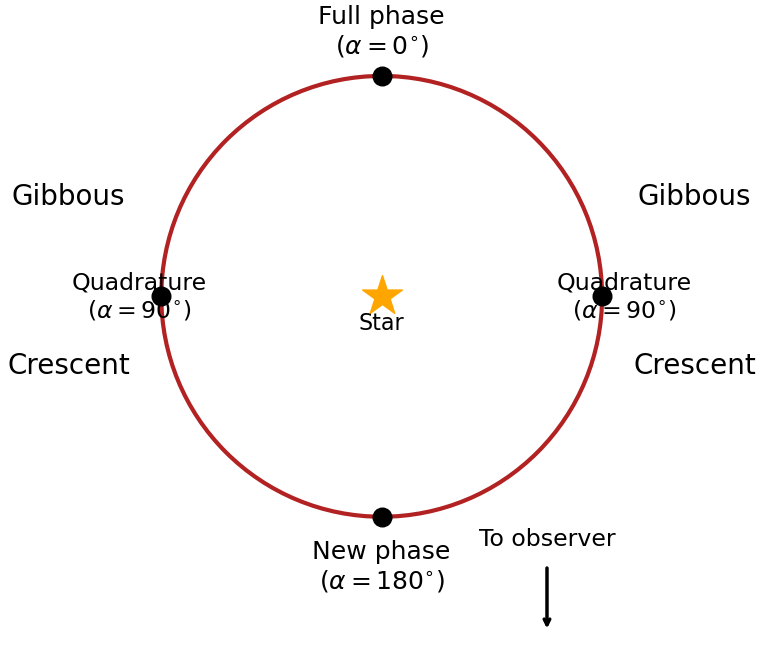}
    \caption{
\emph{Top}: Schematic of the star--planet--observer geometry defining the phase angle $\alpha$, measured between the star--planet and planet--observer directions. The hemisphere of the planet facing the star is fully illuminated (gold) at all times, while the opposite hemisphere remains in shadow. The observer sees only the portion of the illuminated hemisphere that is visible at a given $\alpha$, with full illumination visible only at $\alpha = 0^\circ$.
\emph{Bottom}: Illustration of planetary phases along an edge-on orbit, with the star at the center and the observer direction indicated below the orbit. Although the star-facing hemisphere remains illuminated throughout the orbit, the visible fraction of this hemisphere changes with phase angle. Full phase occurs at $\alpha = 0^\circ$, quadrature at $\alpha = 90^\circ$, and new phase at $\alpha = 180^\circ$, with gibbous and crescent phases spanning the intermediate ranges.}
    \label{fig:phases_geometry}
\end{figure}

When adopted within a yield calculation, we spectrally integrate the wavelength-dependent Earth phase curves across a user-defined band, as specified in the EXOSIMS setup. A V-band transmission function is adopted in all cases below to produce an effective phase curve representative of broadband observations. Within EXOSIMS, the phase function propagates through the full detection and scheduling framework by modifying the planet-star flux ratio and corresponding $\Delta$mag values used in completeness calculations, exposure time estimates, and scheduler decisions. Finally, to aid with understanding later results, Figure~\ref{fig:phases_geometry} depicts the geometry of the phase angle and how phases vary along an orbit viewed edge on.


\subsection{Experimental Setup}
To isolate the impact of phase function choice, we ran two sets of simulations in EXOSIMS: one using the Lambert phase function and one using the Earth-based V-band phase function derived from the spectral Earth model. All other mission and population parameters were held constant. We then compared detection counts, phase-angle distributions, and detection probability curves to quantify how the realistic phase function modifies baseline yield estimates. By incorporating both the Lambert and realistic V-band phase functions, we systematically assessed how different assumptions about phase-dependent brightness influence detection probabilities. This approach quantified the differences in detectability for Earth-like exoplanets and underscored the importance of using realistic phase function data to refine yield models and improve mission strategies for identifying habitable worlds.

In addition to this controlled phase-function comparison, we investigated the combined effects of phase function choice and inner working angle (IWA), a design-critical instrument parameter. The baseline IWA adopted in our simulations corresponds to the nominal instrument configuration listed in Table~\ref{tbl:instrument_params} (0.054 arcsec), which is equivalent to an IWA of $3\,\lambda/D$ at 550nm for a 6m telescope. We performed ensemble simulations spanning IWAs of $2$, $2.5$, $3$, and $3.5\,\lambda/D$ using both the Lambertian and realistic Earth phase functions. For each IWA configuration, we constructed histograms of the phase angles at which detections occurred, enabling us to assess how decreasing IWA expands the accessible range of projected separations and orbital phases, with the resulting phase angle distribution depending on target geometry and scheduler choices.


\section{Results} \label{sec:results}

\begin{figure} [ht!]
    \centering
    \includegraphics[width=0.5\linewidth]{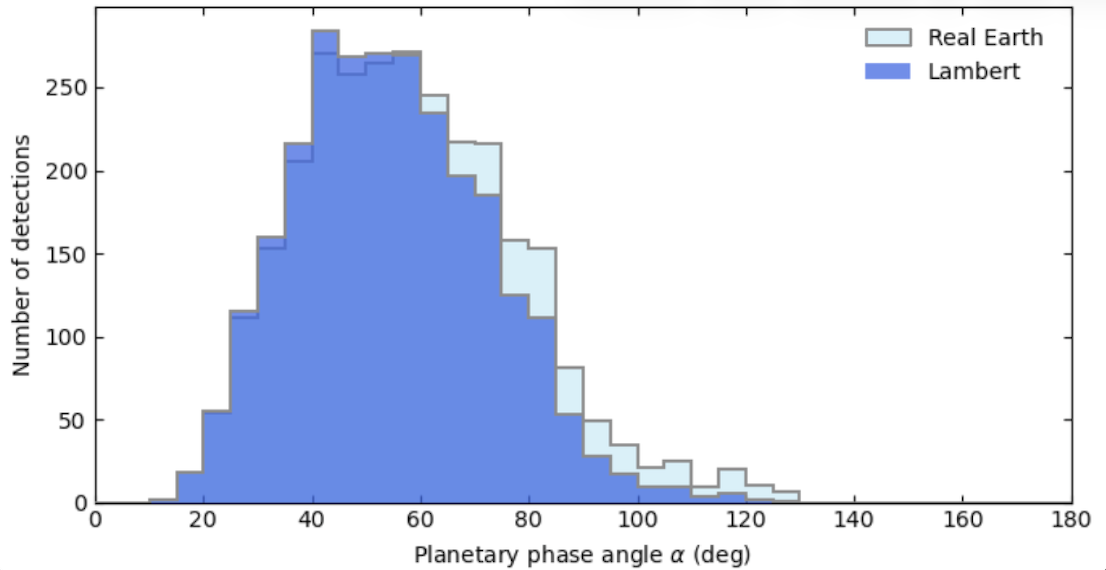}
    \caption{Phase-angle distribution of detections across a 100-seed ensemble for the Lambert and realistic Earth cases, showing that the Lambertian assumption underestimates detections at medium/high phase angles.}
    \label{fig:phaseanglehist}
\end{figure}

We begin by presenting results from the baseline EXOSIMS configurations adopted in this work, including an inner working angle (IWA) of $3\,\lambda/D$ evaluated at V-band. Figure~\ref{fig:phaseanglehist} shows the distribution of planetary phase angles at the time of detection for both the Lambert and realistic Earth phase functions.

The realistic Earth phase function produces a clear redistribution of detections across phase angle. Compared to the Lambert model, detections are shifted toward phase angles where Earth is intrinsically brighter, particularly at medium and high phase angles. This shift reflects the increased brightness of the realistic Earth phase function relative to the Lambert approximation at these phases. While both phase functions produce detections near quadrature, where the brightness models are similar, the realistic phase function enhances detectability at phase angles that are otherwise underrepresented in the Lambert case.

Figure 3 also serves as a validation of the phase-function implementation within EXOSIMS. The realistic Earth phase function is brighter than Lambert at medium and high phase angles (Figure 1), and the corresponding increase in detections occurs at those same phase angles. This behavior demonstrates that the modified phase function propagates through the detection framework in the expected direction and builds confidence in the resulting yield calculations. Having verified that the realistic phase function changes detections preferentially at the phase angles where Earth is brighter than the Lambert model, we next examine whether this redistribution changes the total number of detected planets. We find that using a realistic phase function redistributes the detections towards larger phase angles, producing an ensemble-averaged increase of approximately 20 percent in detection events beyond $90^\circ$ relative to the Lambert phase function.

\begin{figure}[ht!]
\centering
\includegraphics[width=0.5\linewidth]{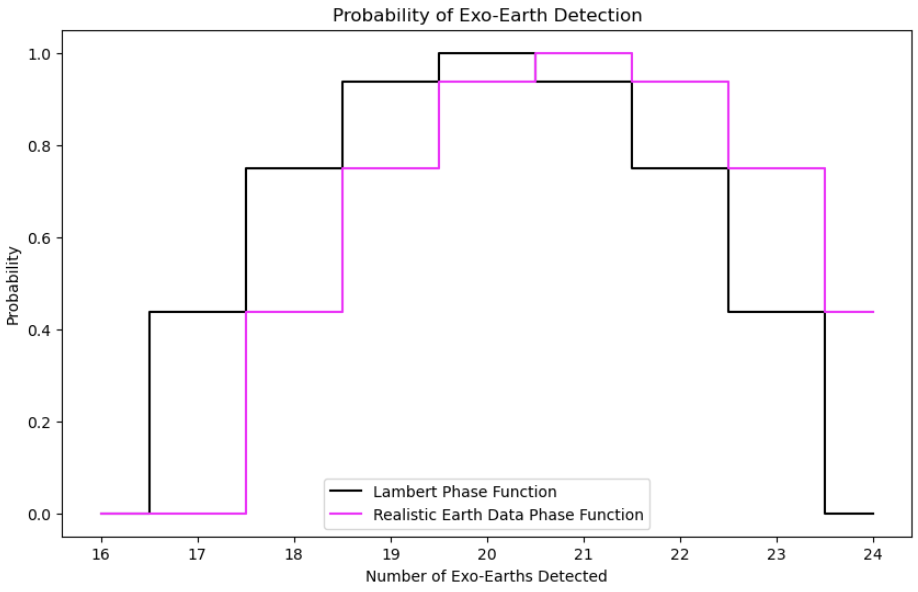}
\caption{Probability distributions for the number of Earth-like exoplanets detected using the Lambert phase function (black) and the realistic Earth phase function (pink). Both cases yield similar median detection counts, with the realistic phase function producing a modest change in the distribution.}
\label{fig:Probability}
\end{figure}

To quantify how these differences impact the total number of detections, Figure~\ref{fig:Probability} illustrates the probability distributions for the number of Earth-like exoplanets detected using the Lambert phase function and the realistic Earth phase function in the baseline setup.

Simulations with the Lambert phase function yielded $20 \pm 2$ detections per mission, while the realistic phase function resulted in an average of $20 \pm 1.5$ detections. This represents an ensemble of 100 Monte Carlo simulation runs. These results show systematic differences in detections but only modest changes in the median yield, indicating that the primary impact is a redistribution across phase angle rather than a large increase in total detections.

\begin{figure} [ht!]
    \centering
    \includegraphics[width=0.5\linewidth]{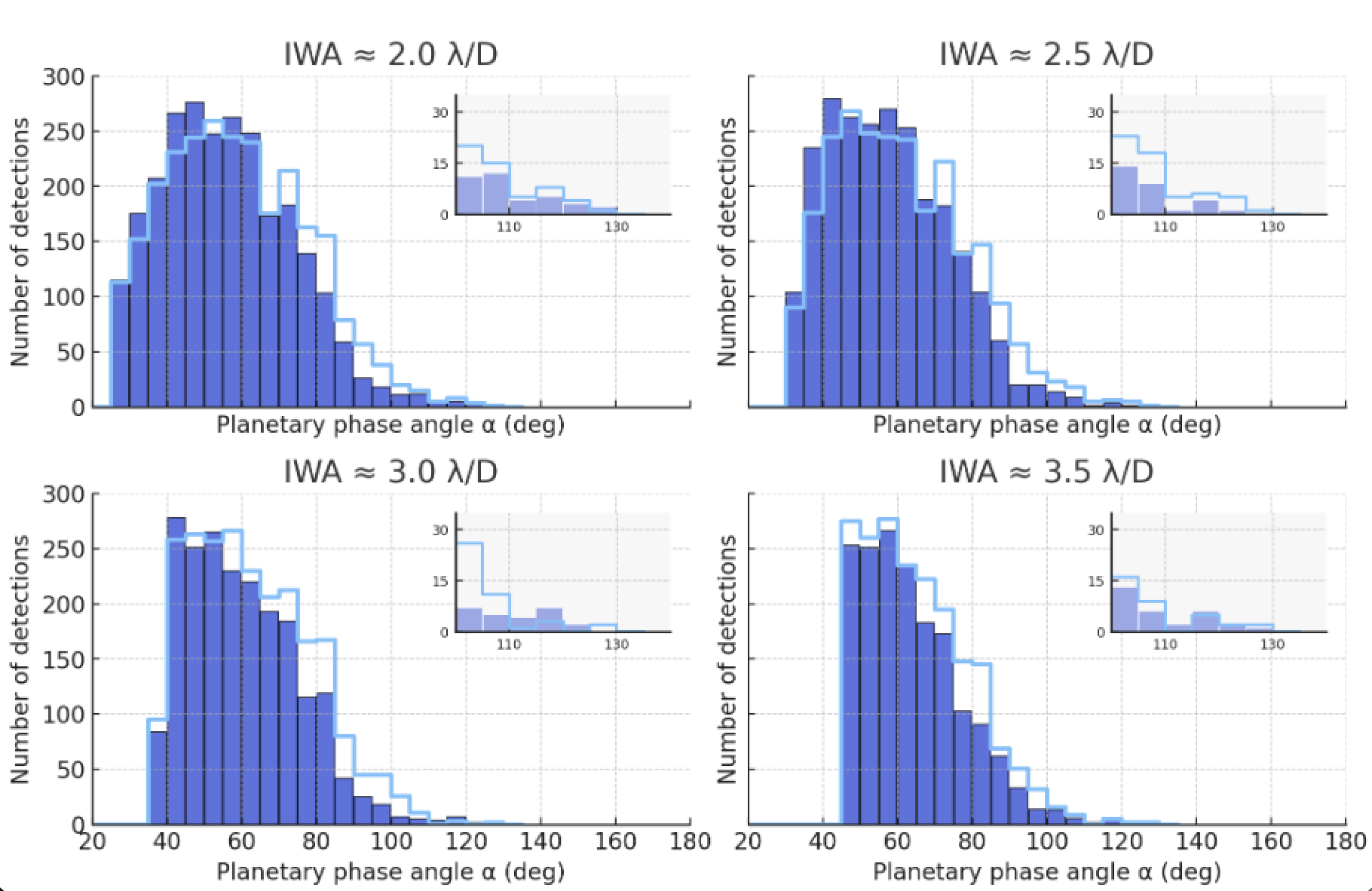}
    \caption{Phase-angle distributions of detections for different coronagraph inner working angles (IWAs). Larger IWAs restrict detections to wider projected separations, limiting the range of accessible orbital phases. As the IWA decreases, the accessible orbital region expands inward toward the star, increasing access to phase angles on both sides of quadrature. Across the ensemble, the realistic Earth phase function produces additional detections at larger phase angles where Earth is intrinsically brighter than the Lambert approximation. The response is not strictly linear because the impact of IWA depends on the projected habitable-zone size of individual systems and on scheduler optimization within the mission simulation. Together, these results demonstrate that IWA and phase-dependent brightness jointly determine which portions of planetary orbits are accessible for detection.}
    \label{fig:IWA}
\end{figure}

Figure~\ref{fig:IWA} shows results from our study that varied the coronagraph IWA. As the IWA decreases, the accessible orbital region expands inward toward the star, allowing detections over a broader range of projected separations and phase angles. This increased access does not exclusively favor crescent phases, but instead enables detections on both sides of quadrature depending on the projected habitable-zone size of a given system. The resulting phase-angle distributions demonstrate that IWA and phase-function assumptions jointly shape the detectability and phase-angle distribution of exo-Earths in an HWO-like mission.

\section{Discussion} \label{sec:discussion}

We examine the detectability of Earth-like exoplanets using the EXOSIMS framework, with a focus on how phase functions impact detection probabilities. By comparing the Lambert phase function with a realistic V-band phase function derived from spectral Earth data, we show that accurately capturing phase-dependent brightness is an important component of direct imaging simulations for future missions designed to characterize potentially habitable worlds.

The realistic Earth phase function produces higher apparent brightness than the Lambert model at medium and high phase angles, especially where scattering from clouds and surface oceans increases Earth’s reflectivity \citep{Robinson2010,zugger2011modeling, vaughan2023chasing}. This leads to a consistent shift in where detections occur. Although the total number of detections does not differ dramatically between the two cases, the realistic phase function redistributes detections toward phase angles where Earth is intrinsically brighter than the Lambert model. Thus, the primary effect of using a realistic phase function is a redistribution of detections across phase angle rather than a large increase in the median number of detected exo-Earths.

This behavior can be understood by considering how exposure time scales with planetary brightness. As shown in Equation~\ref{eq:phase_scaling}  of the Appendix~\ref{sec:appendix}, fractional changes in planetary brightness produce twice that fractional change in exposure time (in the opposite direction). However, over the phase angles where planets are typically sufficiently bright for detection, the fractional differences in \(A_g\,\Phi(\alpha)\) between the Lambert and realistic Earth phase functions are relatively small. As a result, the corresponding changes in exposure time are modest, explaining why the overall yield does not change significantly despite larger differences at more extreme phase angles (Figure~\ref{fig:Probability}).

As shown in Figure~\ref{fig:phaseanglehist}, this redistribution has important implications for follow-up characterization. Enhanced brightness at medium and high phase angles increases detectability in orbital regions where the Lambert model predicts much fainter signals. These detections may shift into crescent phases, where planets are generally fainter and less favorable for immediate spectroscopy. However, detections at these phases can still be valuable, as they help constrain the orbit and enable follow-up observations to be carefully planned, including predicting revisit epochs where spectroscopy would be most efficient. Therefore, realistic phase functions inform not only how many planets a mission may detect, but when they are detected and how that timing aligns with opportunities for atmospheric characterization.

The inner working angle plays an essential role in moderating the impact of the phase function. As shown in Figure~\ref{fig:phaseanglehist}, detections are concentrated at intermediate phase angles, where there is a favorable balance between illumination and angular separation. In (Figure~\ref{fig:IWA}), smaller IWAs expand the range of projected separations accessible to the coronagraph. For individual systems, this can expose both gibbous and crescent orbital configurations; across the full target ensemble, the effect depends on stellar distance, habitable-zone angular size, and scheduler choices. The realistic Earth phase function then has the strongest impact where those newly accessible phases overlap with phase angles at which Earth is brighter than the Lambert model. 

When the IWA is large, the angular region where the realistic phase function differs most strongly from the Lambert model is inaccessible, suppressing the effect of the realistic brightness model and favoring detections at larger phase angles, including gibbous configurations (Figure~\ref{fig:IWA}). When the IWA is smaller,these differences become more pronounced because a broader range of orbital phases becomes accessible, allowing the realistic Earth phase function to affect detections over a larger portion of the orbit.

These findings are consistent with previous yield studies \citep{Stark2014,Morgan2017}, which show that improved brightness models and observing assumptions tend to modify the distribution of detections more strongly than the total yield. In particular, these studies demonstrate that while changes to planet brightness or observing strategy can shift when and where detections occur, the overall number of detected exoplanets may only vary on the order of tens of percent under realistic mission constraints. Observational constraints, particularly the IWA, limit the orbital range accessible to the telescope and therefore limit how much brightness differences at large phase angles can change the overall yield. Regardless, realistic phase functions provide a more accurate representation of when planets are detectable and can improve how yield tools capture planetary visibility and follow-up potential.

Despite these advancements, our simulations rely on the main simplifying assumption that all of the targets are like Modern Earth. However, the true population of small planets is expected to be diverse, spanning Earth-sized worlds through sub-Neptunes \citep{fulton2017california, bean2021nature}. Even Earth itself has exhibited significant spectral variability over geologic time \citep{krissansen2018disequilibrium, arney2016pale}. Thus, the real use case of HWO may be more complicated. All habitable-zone rocky planets were treated as Earth twins, even though real exo-Earths may differ in atmospheric composition, cloud cover, climate state, or surface properties. The realistic Earth phase function used here represents only one planetary environment. Future refinements could incorporate multiple phase functions representing a broader diversity of terrestrial worlds to better capture the full range of possible outcomes.

In future work, more sophisticated phase-function models and phase-dependent spectral information can be incorporated into EXOSIMS to further refine detectability predictions. Integrating these developments into observational mission planning can help optimize scheduling strategies, coronagraph design decisions, and characterization priorities. By increasing the realism of detection modeling within EXOSIMS, this work provides practical insights for optimizing future missions aimed at identifying and studying potentially habitable worlds.

Overall, this work demonstrates that realistic phase functions play a significant role in shaping when and where detections occur, even when the total number of detections remains similar. By improving the physical realism of direct-imaging simulations, this work supports the broader goal of preparing HWO to robustly detect and characterize Earth-like exoplanets.

\section{Conclusions} \label{sec:conclusions}

In this work, we implemented a realistic, Earth-like phase function into EXOSIMS and compared its impact on detection yield and phase-angle distributions relative to the default Lambert phase function. We then quantified how phase-dependent brightness and inner working angle jointly affect detectability for Earth-like planets. We emphasize the most important conclusions below:

\begin{itemize}
\item Implementing realistic planetary phase functions, especially for Earth-like worlds, is essential for understanding planetary reflectivity and its impact on detectability in direct imaging missions.
\item The Lambert phase function assumes isotropic reflection, providing a baseline for detectability, while realistic phase functions account for phase-dependent atmospheric and surface contributions to reflectivity.
\item Simulations using the Lambert phase function show a broad distribution of detections near quadrature, reflecting the geometry of detection.
\item Introducing realistic Earth phase functions redistributes detections toward medium/high phase angles where Earth is brighter than Lambert
\item Detection distributions highlight the differences between the Lambert and realistic phase functions, with realistic models enhancing detectability at smaller separations.
\item An IWA sweep with both phase functions demonstrates that smaller coronagraph inner working angles expand accessible orbital phase space, illustrating that IWA and phase-function assumptions jointly shape the detectability of Earth-like planets in HWO-like architectures.
\item These results motivate the adoption of realistic phase functions and IWA analysis, as well as further explanation of other important mission parameters. These represent a step towards making yield models more physically grounded and science forward for direct imaging missions. 
\end{itemize}

\section{Acknowledgments} \label{sec:acknowledgments}
SF gratefully acknowledges support from the Jet Propulsion Laboratory Strategic University Research Partnerships (SURP) Program and the University of Arizona through the State of Arizona's Technology and Research Initiative Fund. SF would also like to thank her HABLab group members and collaborators at NASA JPL for their support.
\clearpage


\bibliographystyle{plainnat} 
\bibliography{Footeexport} 

\section{Appendix} \label{sec:appendix}
\textbf{\section*{Approximating the Sensitivity of Integration Time to Phase Function Value}}

From Equation (6) in the direct imaging instrument noise model described in \citet{robinson2016coronagraph}, the exposure time ($\Delta t$) required to achieve a specified SNR is given by,
\begin{equation}
    \Delta t = \frac{c_{\rm p} + 2 c_{\rm b}}{c_{\rm p}^2} {\rm SNR}^2 \ ,
\end{equation}
where $c_{\rm p}$ and $c_{\rm b}$ are the planet and background count rates, respectively. The challenging nature of exo-Earth direct imaging often implies that these sources are fainter than the background, yielding the simplification,
\begin{equation}
    \Delta t \approx \frac{ 2 c_{\rm b}}{c_{\rm p}^2} {\rm SNR}^2 \ .
\end{equation}
To explore sensitivity to the planetary signal, we differentiate this expression, yielding,
\begin{equation}
    \frac{\partial \Delta t}{\partial c_{\rm p}} \approx - \frac{ 4 c_{\rm b}}{c_{\rm p}^3} {\rm SNR}^2 \ ,
\end{equation}
or,
\begin{equation}
    \frac{\partial \ln \Delta t}{\partial \ln c_{\rm p}} \approx -2 .
\end{equation}
As the planet brightness scales with the product of the geometric albedo and the planetary phase function, we have,
\begin{equation}
    c_{\rm p} \propto A_{\rm g} \Phi(\alpha) \,
\end{equation}
so that, as a differential, we have,
\begin{equation}
    \delta \ln c_{\rm p} = \delta \ln A_{\rm g} \Phi \ .
\end{equation}
Combining this with our earlier derivative, we see that,
\begin{equation}
    \delta \ln \Delta t \approx -2 \delta \ln A_{\rm g} \Phi \ .
    \label{eq:phase_scaling}
\end{equation}
Thus, this analytic scaling suggests that for an increase in the planetary brightness, we expect that the fractional change in the exposure time required to achieve a given SNR on that target decreases by a factor of about twice the increase in the relative brightness.

\end{document}